\documentclass[submission,copyright,creativecommons]{eptcs}
\providecommand{\event}{SNR'2021}

\usepackage{microtype}

\usepackage{float}
\usepackage{rwthcolors}
\usepackage{amsmath}
\usepackage{amssymb,bm}
\usepackage{subfig}
\usepackage{hyperref}
\usepackage[]{cleveref} 
\usepackage{booktabs}
\usepackage{tabularx}
\usepackage{epstopdf}
\usepackage{graphicx}
\usepackage{tikz}
\usetikzlibrary{calc,shapes,arrows,automata}
\usetikzlibrary{shapes,arrows,decorations.markings,decorations.pathmorphing}
\usetikzlibrary{decorations.pathreplacing}
\usepackage{multirow}
\usepackage{xspace}
\usepackage{wrapfig}
\usepackage{todonotes}
\usepackage[]{algorithm2e}
\usepackage{tikz}
\usepackage{url}
\usepackage{doi}
\usetikzlibrary{shapes,arrows}

\graphicspath{{./pics/}}

\newcommand{\ie}{i.e.\xspace}

\newcommand{\eg}{e.g.\xspace}

\newcommand{\R}{\ensuremath{\mathbb{R}}}
\renewcommand{\H}{\ensuremath{\mathcal{H}}}
\newcommand{\Flow}{\mathit{Flow}}
\newcommand{\Pred}[1]{\mathit{Pred}_{#1}}

\newcommand{\Loc}{\ensuremath{\mathit{Loc}}\xspace}
\newcommand{\Var}{\ensuremath{\mathit{Var}}\xspace}

\newcommand{\Edge}{\ensuremath{\mathit{Edge}}\xspace}

\newcommand{\Inv}{\ensuremath{\mathit{Inv}}\xspace}
\newcommand{\Init}{\ensuremath{\mathit{Init}}\xspace}
\newcommand{\hypro}{\textsc{HyPro}\xspace}

\newcommand{\flowstar}{\textsc{Flow*}\xspace}

\newcommand{\spaceex}{\textsc{SpaceEx}\xspace}

\newcommand{\hycreate}{\textsc{HyCreate}\xspace}

\newcommand{\cora}{\textsc{Cora}\xspace}

\newcommand{\argos}{\textsc{ARGoS}\xspace}

\newcommand{\parrep}[1]{\noindent\textit{\textbf{#1}}}

\newcommand{\shd}{\texttt{shd I}\xspace}
\newcommand{\shdOpt}{\texttt{shd II}\xspace}
\newcommand{\lsync}{\texttt{lsync I}\xspace}
\newcommand{\lsyncOpt}{\texttt{lsync II}\xspace}

\newtheorem{definition}{Definition}

\title{Robot Swarms as Hybrid Systems: Modelling and Verification}
\def\titlerunning{Robot Swarms as Hybrid Systems: Modelling and Verification}

\author{Stefan Schupp\institute{TU Wien, Vienna, Austria 
	} \and
Francesco Leofante\institute{Imperial College London, London, United Kingdom 
} \and
Leander Behr\institute{RWTH Aachen University, Aachen, Germany} \and
Erika Ábrahám\institute{RWTH Aachen University, Aachen, Germany} \and
Armando Taccella\institute{University of Genoa, Genoa, Italy}}
\def\authorrunning{S. Schupp, F. Leofante, L. Behr, E. Ábrahám, A. Taccella}

\begin{document}

\maketitle

\begin{abstract}
 A swarm robotic system consists of a team of robots performing
 cooperative tasks without any centralized coordination. In principle,
 swarms enable flexible and scalable solutions; however, designing
 individual control algorithms that can guarantee a required global
 behavior is difficult. Formal methods have been
 suggested by several researchers as a mean to increase confidence in the
 behavior of the swarm. In this work, we propose to model swarms as hybrid systems and use reachability analysis to
verify their properties. We discuss challenges and report on the experience
gained from applying hybrid formalisms to the verification of a swarm robotic
system.
\end{abstract}

\section{Introduction}
\label{sec:intro}
Swarm robotic systems are distributed systems wherein a set of robots 
cooperatively perform a
task, without any centralized coordination~\cite{DBLP:conf/dars/Parker00}.
Although individual robots are governed by relatively simple reactive 
controllers, interactions within the swarm may give rise to complex behaviors 
that were not explicitly programmed. Ultimately, these behaviors enable the 
swarm to achieve goals that would
defy each single robot in isolation, or would require more expensive
robots to achieve the same goals as effectively as the swarm does -- see,
e.g,~\cite{DBLP:journals/swarm/BrambillaFBD13,DBLP:conf/sab/Sahin04} for some 
examples.

While understanding individual robot behavior is easy, predicting the
overall swarm behavior is difficult, and thus engineering controllers
for individual robots that will guarantee a desired swarm behavior is
not a straightforward task. Traditionally, the analysis of swarms
is carried out either by testing real robot implementations,
or by computational 
simulations~\cite{DBLP:conf/bioadit/LabellaDD04,DBLP:conf/sab/LiuWSCD06}; 
however, these approaches provide little guarantees as they suffer from 
intrinsically incomplete coverage.
As suggested by many authors, higher levels of assurance in swarm behavior can 
be obtained via formal 
methods~\cite{DBLP:conf/sefm/RouffVHTR04,DBLP:journals/sosym/PenaRHC11,DBLP:journals/ras/KonurDF12,DBLP:conf/aamas/BrambillaPBD12,DBLP:conf/ijcai/KouvarosL15,DBLP:conf/ijcai/LomuscioP18}.
 However most approaches abstract
away details about the continuous dynamics of the robots, which may
indeed be crucial for the emergence of desired behaviors.

In this paper, we fill this gap by investigating the problem of
providing assurance guarantees for swarm robotic systems through mixed discrete-continuous
formalisms. In particular, we propose to use hybrid 
automata~\cite{henzinger:hybrid} to
define richer models of swarms and employ tools for reachability
analysis to overcome the incomplete coverage of
testing and simulation. Here we report on the application of
reachability analysis of hybrid automata to verify a controller developed for
swarm applications. Modeling challenges are discussed, together with
features and limitations of tools for reachability analysis. 
Our contributions can be summarized as follows:
\begin{itemize}
\item Applying hybrid automata modeling and reachability analysis as a
  method to formally engineer robot swarms; to the extent of our
  knowledge, this is the first contribution in this direction.
\item Modeling and analyzing a simple but realistic swarm robotic
  system entailing synchronization without central
  coordination.
 \item Discussing challenges together with new solutions proposed to 
 overcome modeling and scalability issues.

\end{itemize}

 The remainder of this paper is organized as
 follows. \Cref{sec:back} introduces background notions on swarm systems and 
 their verification. \Cref{sec:ha} contextualizes the analysis of robot
 swarms as hybrid systems and \Cref{sec:results} presents our case study, 
 together with experimental results. Finally, \Cref{sec:future} provides 
 concluding  remarks and future directions of research.

\section{Background}
\label{sec:back}

\subsection{Swarm Robotic Systems}
\label{ssec:swarm}
 A robot swarm can be
 defined as a specific kind of distributed autonomous mobile robotic
 system wherein a set of robots is meant to perform some kind of
 collective task~\cite{DBLP:conf/dars/Parker00}. This is achieved following 
 decentralized, behavior-based control
 rules relying on ``social'' interactions among the robots.

A swarm can be characterized as
in~\cite{DBLP:journals/swarm/BrambillaFBD13}: robots are autonomous;
they are situated in an environment and act to modify it; robot's
sensing and communication capabilities are local; robots do not have
access to centralized control and/or global knowledge and
cooperate to accomplish a task.
The cooperation among robots happens in different ways,
including explicit communication through,
e.g., a wireless network, or implicitly through \emph{stigmergy}~\cite{beckers2000fom},
i.e., robots sense changes made by
other robots, and then adjust their behavior accordingly. In any case, the 
collective behavior is
not predefined to any extent at the global level, but it is most
likely to be \emph{emergent}, i.e., the result of several
local robot-to-robot and robot-to-environment interactions.
While the definition of ``emergence'' has been the subject of different
contributions, here we take as working definition the one given
in~\cite{mataric1993designing} asserting that emergent behaviors are
characterized by two properties: $(i)$ they are manifested
by global states or time-extended patterns which are not explicitly
programmed in, but result from local interactions among a system's
components; $(ii)$ they are considered interesting based on
some observer-established metric.
%
%

In this work we study emergent behaviors within swarms of
MarXbots~\cite{DBLP:conf/iros/BonaniLMRBRVBM10},
mobile robots that have been conceived and built through several European 
projects focusing on swarm  -- see,
e.g,~\cite{DBLP:journals/arobots/MondadaPGKFDNGD04}; successful
applications of such robots can be found, e.g.,
in~\cite{DBLP:conf/dars/FerranteBBD10}
and~\cite{DBLP:journals/swarm/DucatelleCPG11}.
To support our experimentation, we use
\argos\cite{DBLP:journals/swarm/PinciroliTOPBBMFCDBGD12}, a simulator
designed to efficiently simulate complex experiments involving large
swarms of robots, developed within the same European projects mentioned above.

\subsection{Formal Verification of Swarms}
\label{ssec:fm}
%
To
the best of our knowledge, the first contribution along this line of
investigation is~\cite{DBLP:conf/sefm/RouffVHTR04}, wherein the
authors investigated the applicability of formal methods to the
verification and validation of spacecraft using swarm technology. More
specifically, they considered a number of approaches including
Communicating Sequential Processes (CSP), process algebras, X-machines
and Unity Logic. However, their conclusion at the time of the
contribution (2004) was that none of the  approaches had all
the properties required to assure correct behavior and interactions of
swarms in the context of the ANTS (Autonomous Nano Technology
Swarm) concept mission. In~\cite{DBLP:journals/sosym/PenaRHC11},
agent-oriented software engineering is investigated to provide
support for the development of swarm robotics systems, still in the context of
the ANTS mission. However, no mention related to assuring behaviors is
to be found in the contribution which is largely confined to
model-based design and implementation techniques. A series of papers
by Dixon et al. -- see,
e.g.,~\cite{DBLP:journals/ras/KonurDF12} for the most recent
contribution in the series --
explores the potential of modeling robot swarms using a composition of
(probabilistic) finite state machines and of proving swarm-level
requirements using model checking of (probabilistic) temporal logic.
Noticeably, in the case of probabilistic models and logic, the model
of each single robot controller is very close to the actual
implementation, and model checking of relevant properties is reported
to be feasible for swarms of relatively small size -- less than 4 robots
according to the experiments
in~\cite{DBLP:journals/ras/KonurDF12}. 

The problem of verifying swarm systems and their emergent properties was also 
considered more recently 
in~\cite{DBLP:conf/ijcai/KouvarosL15,DBLP:conf/aaai/KouvarosL15,DBLP:conf/ijcai/LomuscioP18,DBLP:conf/aamas/LomuscioP19},
where a number of techniques have been proposed to improve the scalability of 
existing verification algorithms. However, these
approaches abstract away the physical dynamics of the robots of the environment,
which may be determinant for the emergence of unforeseen behaviors. In 
robotics, this is more than
just an abstract principle, because the implements
are physical agents that can damage the environment, and thus should
be subject to stringent requirements -- see, e.g.,  ISO/TC 299
published standards about safety requirements for various kinds of
robots. Collective behaviors may sometimes emerge not only from the current 
state of the single robots, but also from the continuous dynamics in time and
space. In order to model such complex properties, discrete transitions as well 
as continuous dynamics must be included in the formal model. For this reason, in this work we 
elaborate on the choice of \emph{hybrid automata} as a modeling formalism.

\section{Hybrid Systems Reachability Analysis}
\label{sec:ha}
The approaches discussed in the previous section abstract the physical
components of robots and simplify their interaction with the
environment. Indeed, all robotic systems are a combination of
programmed digital controllers and implements -- sensors and
actuators -- interfacing with the physical world. Therefore, an
accurate model of robot operation should include both the (discrete)
control states and the (continuously) varying physical quantities. In
the case of robot swarms, we believe that failing to take into account
the physical components, may hamper our ability
to determine whether the swarm will behave correctly
at all times. \emph{Hybrid automata} are a well-established formalism
to model systems combining discrete states and continuously varying
quantities, and methods for their verification have been successfully
developed.

\subsection{Hybrid Automata}


\begin{definition}[Hybrid automaton: Syntax~\cite{henzinger:hybrid}]
	\label{def:hybrid_automaton}
\!A \emph{hybrid automaton} $\H=(\Loc{,}\Var{,}\Flow{,}\Inv{,}\Edge{,}\Init)$ is a tuple consisting of:
\begin{itemize}
\item A finite set \Loc of \emph{locations} or \emph{control modes}.
\item A finite ordered set $\Var=\{x_1,\ldots,x_n\}$ of real-valued
  \emph{variables}; we also use the vector notation $\vec{x}=(x_1,\ldots,x_n)$.
  The number $n$ is called the \emph{dimension} of $\H$. By $\dot{\Var}$ we
denote the set $\{\dot{x}_1,\ldots,\dot{x}_n\}$ of dotted variables (which
represent first derivatives during continuous
change), and by $\Var'$ the set $\{x_1',\ldots,x_n'\}$ of primed variables
(which represent values directly after a discrete change). Furthermore,
$\Pred{X}$ is the set of all predicates with free variables from $X$.
\item $\Flow:\Loc\rightarrow\Pred{\Var\cup\dot{\Var}}$ specifies for each
  location its \emph{flow} or \emph{dynamics}.
\item $\Inv:\Loc\rightarrow\Pred{\Var}$ assigns to each location an
  \emph{invariant}.
\item $\Edge \subseteq \Loc\times
  \Pred{\Var}\times\Pred{\Var\cup\Var'}\times\Loc$ is a finite set of
  \emph{discrete transitions} or \emph{jumps}. For a jump
  $(l_1,g,r,l_2)\in\Edge$, $l_1$ is its \emph{source} location, $l_2$ is its
  \emph{target} location, $g$ specifies the jump's \emph{guard}, and $r$ its
  \emph{reset} function, where primed variables represent the state after the
  step.
\item $\Init:\Loc\rightarrow\Pred{\Var}$ assigns to each location an
  \emph{initial} predicate.
\end{itemize}
\end{definition}

\begin{figure}
 	\centering
 	\scalebox{0.8}{\begin{tikzpicture}[node distance = 2cm, auto]
\tikzstyle{cloud} = [draw, rectangle, rounded corners, node distance=3cm,
minimum height=2em]
\tikzstyle{line} = [draw, -latex']

\node [cloud, ] (lo) {\begin{tabular}{c}
	\texttt{driving}\\
	$\dot{x} = v \cos(\theta)$\\
	$\dot{y} = v \sin(\theta)$\\
	$\dot{\theta} = w$\\
	$\dot{w} = 0 $\\
	$\dot{clock} = 1$
	\\
	$clock \leq 0.1$
	\end{tabular}};
\node [left of=lo, node distance=3.5cm] (null) {\begin{tabular}{l}
	$x = 0 \wedge y = 0 \wedge$\\
	$\theta = 0 \wedge w = 1$\\
	$clock = 0$
	\end{tabular}};

\path [line] (null) -- (lo);
\path [line] (lo) edge [out=30,in=330,looseness=5] node[right] {\begin{tabular}{l}
	$clock \geq 0.1$\\
	$w' := -w$\\
	$clock' := 0$
	\end{tabular}	} (lo);

\end{tikzpicture}}
 	\caption{Hybrid automaton for a differential drive robot with constant speed $v=1$.}
 	\label{fig:HADiffDrive}
 \end{figure}
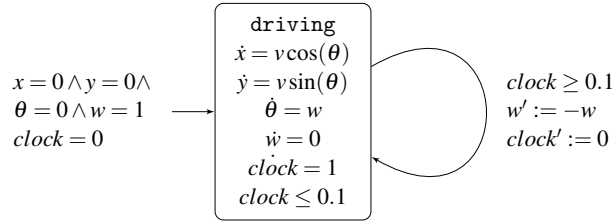

 An example hybrid automaton that models a differential drive robot driving a chicane is depicted in \Cref{fig:HADiffDrive}.
 The system has $5$ variables used for describing the relevant physical
 quantities of the robot (position, steering and angular velocity). An initial
 variable assignment
 (\emph{initial set}) specifies the initial conditions of the system. The
 automaton has one \emph{control mode} (\texttt{driving}) modeling
 the continuous
 behavior of the robot which is specified by ordinary differential equations (ODEs). The continuous behavior inside the location is limited
 by the invariant on the variable \emph{clock}. This variable models a timer, which guards the only transition in the model. As
 soon as the guard condition $clock \geq 0.1$ is satisfied, the transition is \emph{enabled} and can be taken.

The \emph{state} of a hybrid automaton is specified by a pair $(l,\vec{v})\in\Loc\times\R^n$ of a location and a variable valuation. The evolution of a hybrid system over time can be described by a run in the respective hybrid automaton, which includes both continuous
evolution (\emph{flow}) and discrete state changes (\emph{jump}). The formal semantics of a hybrid automaton is defined as follows.

\begin{definition}[Hybrid automaton: Semantics]
The operational semantics of a  hybrid automaton $\H =
(\Loc,\Var,\Flow,\Inv,\Edge,\Init)$ of dimension $n$ is defined by the
following rules:
	\[
\begin{array}[h]{rl}
	\begin{array}[h]{c}
		l\in\Loc\quad \vec{v},\vec{v}'\in\R^n\\
f:[0,\delta]\rightarrow \R^n \quad \mathit{df}/\mathit{dt}=\dot{f}:(0,\delta)\rightarrow\R^n\quad

f(0)=\vec{v}\quad f(\delta)=\vec{v}'\\
\forall \epsilon\in(0,\delta).\
f(\epsilon),\dot{f}(\epsilon)\models\Flow(l)\quad
\forall \epsilon\in[0,\delta].\ f(\epsilon)\models\Inv(l)
\\\hline
		(l,\vec{v})\stackrel{\delta}{\rightarrow} (l,\vec{v}')
	\end{array}
	& \begin{array}{c}\\ \\
	\texttt{Rule}_{\ \texttt{flow}}
\end{array}
\end{array}
\]
\vspace*{2ex}
\[
\begin{array}[h]{rl}
	\begin{array}[h]{c}
		e=(l,g,r,l')\in\Edge\quad \vec{v},\vec{v}'\in\R^n\quad \vec{v}\models g \quad
\vec{v},\vec{v}'\models r\quad \vec{v}'\models\Inv(l')\\\hline
		(l,\vec{v})\stackrel{e}{\rightarrow} (l',\vec{v}')
	\end{array}
	&
	\texttt{Rule}_{\ \texttt{jump}}
\end{array}
\]
A \emph{path} of $\H$ is a (finite or infinite) sequence
$(l_0,\vec{v}_0) \stackrel{\delta_0}{\rightarrow} (l_1,\vec{v}_1)\stackrel{e_1}
{\rightarrow}(l_2,\vec{v}_2)\stackrel{\delta_2}{\rightarrow}(l_3,\vec{v}_3
)\stackrel{e_3}{\rightarrow}(l_4,\vec{v}_4)\stackrel{\delta_4}{\rightarrow
}\ldots$ with $(l_i,\vec{v}_i)$ states of $H$, $\delta_i\in\R_{\geq 0}$,
$e_i\in\Edge$, and $\vec{v}_0\models\Init(l_0)\wedge \Inv(l_0)$. A state
$(l,\vec{v})$ is \emph{reachable} in $\H$ if there is a path
$(l_0,\vec{v}_0)\stackrel{\delta_0}{\rightarrow}(l_1,\vec{v}_1)\stackrel{e_1}
{\rightarrow}(l_2,\vec{v}_2)\stackrel{\delta_2}{\rightarrow}\ldots$ of $\H$
with $(l,\vec{v})=(l_i,\vec{v}_i)$ for some $i\geq 0$.
\end{definition}

Paths defined over states naturally generalize
to paths over state sets $(l,V)=\{(l,\vec{v})\,|\, \vec{v}\in V\}$ for $V\subseteq\R^n$: a \emph{symbolic path}
$(l_0,V_0) \stackrel{\delta_0}{\rightarrow} (l_1,V_1)\stackrel{e_1}
{\rightarrow}(l_2,V_2)\ldots$ represents the set of all paths
$(l_0,\vec{v}_0) \stackrel{\delta_0}{\rightarrow} (l_1,\vec{v}_1)\stackrel{e_1}
{\rightarrow}(l_2,\vec{v}_2)\ldots$ with $\vec{v}_i\in V_i$ for all $i$.

The above formalism can be extended with \emph{urgent} jumps, which force the control to take a jump as soon as an urgent jump is enabled. In graphical illustrations, we mark urgent jumps with a star. In our models, in the source locations of urgent jumps no time can pass, thus urgency could be modeled by introducing a fresh clock with derivative $1$, resetting it to 0 when entering the location, and adding an invariant stating that the variable value is 0. However, some approaches can analyse reachability more efficiently when using urgent jumps.

To model compositional hybrid systems involving
communication between the components, we use the \emph{parallel composition} of hybrid automata. Besides shared variables, we can model communication by annotating jumps with
\emph{synchronization labels}. Semantically, time evolves in all components in parallel; a jump of a component can be taken only if all components that have the given label take a jump with that label simultaneously.
Thus when building the parallel composition syntactically as a single hybrid automaton, a composed jump needs to be added for each possible combination of synchronizing jumps. Therefore, in general, the size of the composition increases exponentially not only in the number of locations, but also in the number of jumps.

\subsection{Reachability Analysis}

Once a hybrid automaton model of a given hybrid system has been formalized, we are interested in analyzing its behavior. Given that the
hybrid systems we are considering are physical agents that can act and modify the environment, we  want to enforce stringent safety
requirements upon their behavior. Such requirements can be formalized as sets of states to be avoided in the state space of a hybrid
automaton.

The \emph{reachability problem} for hybrid automata is the problem to
decide whether a given state (or any state from a given set) is
reachable in a hybrid automaton. As the reachability problem for
hybrid automata is in general undecidable, some approaches aim at
computing an \emph{over-approximation} of the set of reachable states
of a given hybrid automaton. We focus on approaches based on
\emph{flowpipe-construction}, which iteratively over-approximate the
set of reachable states by the union of a set of state sets -- see \eg \cite{DBLP:conf/syde/Frehse15} for further details. To
\emph{represent} a state set, typically either a \emph{geometric} or a
\emph{symbolic} representation is used. Geometric representations
specify state sets by geometric objects like boxes~\cite{moore2009introduction}, (convex)
polytopes~\cite{ziegler1995lectures}, template polyhedra~\cite{10.1007/978-3-540-78800-3_14}, zonotopes~\cite{DBLP:confhybridGirard05}, or ellipsoids~\cite{ellipsoids}, whereas symbolic
representations
use, e.g., support functions~\cite{LeGuernicG10} or Taylor models~\cite{DBLP:conf/rtss/ChenAS12}. These representations
might have major differences in the precision of the representation
(the size of over-approximation), the memory requirements and the
computational effort needed to apply operations like intersection,
union, linear transformation, Minkowski sum or test for emptiness.

For a given initial state set $p$, flowpipe-construction-based approaches
first over-approximate the set of states reachable from $p$ via time evolution.
Time evolution is usually restricted to a
\emph{time horizon} (either per location or for the whole execution),
which is divided into smaller time segments. The time successors (called the \emph{flowpipe}) from
$p$ are over-approximated by a sequence of state sets $p_1,\ldots,p_k$ (\emph{flowpipe segments}), one for each time segment. For each of these, all possible jump successors are computed and for each jump successor set the whole procedure is repeated iteratively. The algorithm terminates if either a fixed point is detected, or the time horizon is reached, or a maximal number of jumps (\emph{jump depth}) have been considered,
or if all successor sets are empty (i.e. there are no jump successors from the flowpipes). The union of all computed state sets over-approximates reachability within the given time and jump bounds.
A non-exhaustive list of software implementations of flowpipe-construction-based methods includes
\cora\cite{cora}, \flowstar\cite{flowstar}, \hycreate\cite{hycreate}, \hypro\cite{hypro} and \spaceex\cite{spaceex}.

The resulting over-approximations for time- and jump-successor states for all reachable paths (over state sets) can be collected in the \emph{reachability tree} in which nodes represent time-evolution and the parent-child relation between nodes represents a discrete jump.
Note that since the method computes over-approximations of sets of reachable states, consequently only over-approximations of paths over state sets are computed and stored in the reachability tree.
Thus it may happen, that the reachability tree contains sub-trees which are not reachable in the hybrid system model, but are only included due to the over-approximation.
For further details on the reachability tree we refer to~\cite{Schupp:767529}.

\section{Case Study}
\label{sec:results}

This section presents our analysis pertaining the application of hybrid systems
verification techniques to a case study in swarm robotics. In the following we
introduce the problem considered, together
with a discussion on modeling and verification in the hybrid setting.

\subsection{Synchronization without Central Coordination}
\begin{figure}[t]
	\begin{center}
		\subfloat[  ]
		{
			\includegraphics[width=0.35\textwidth]{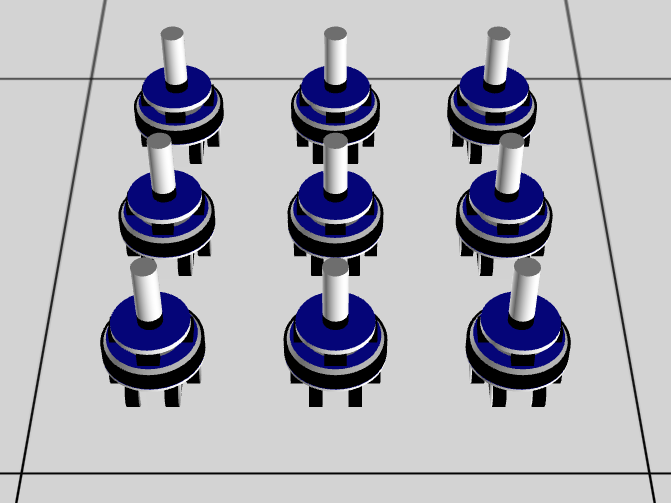}
			\label{fig:start1}
		}
		\subfloat[ ]
		{
			\includegraphics[width=0.35\textwidth]{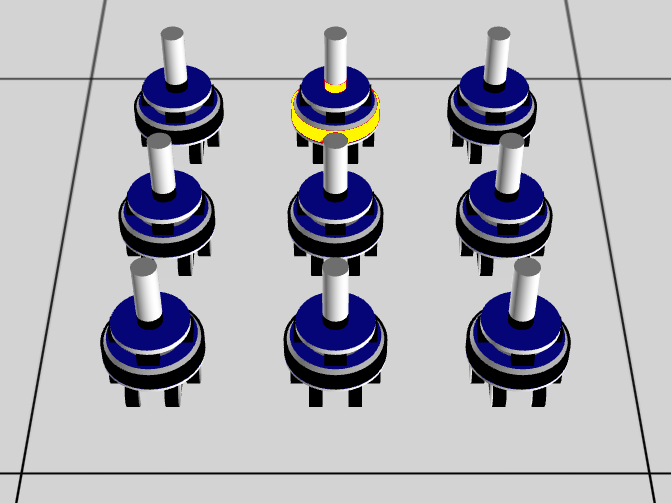}
			\label{fig:step1}
		}
		\\
		\subfloat[ ]
		{
			\includegraphics[width=0.35\textwidth]{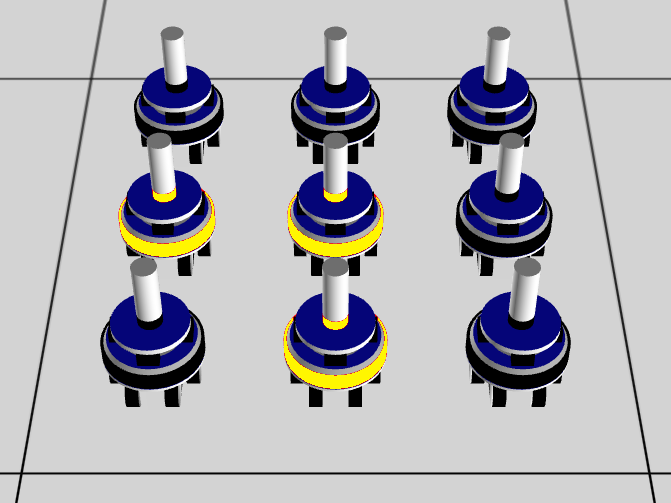}
			\label{fig:step2}
		}
		\subfloat[ ]
		{
			\includegraphics[width=0.35\textwidth]{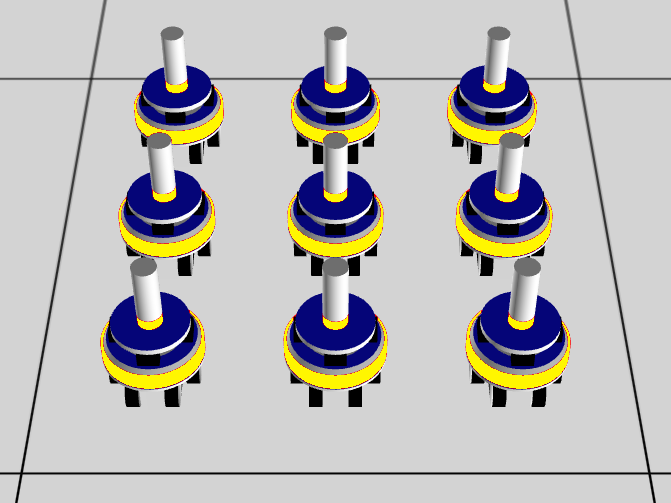}
			\label{fig:end}
		}
	\end{center}
	\caption{\label{fig:sync} Synchronization in the simulator.}

\end{figure}

\paragraph*{Problem description.} Mutual synchronization is a natural phenomenon
whereby a population of individuals synchronize over a common behavior and act
in perfect unison among themselves --
see,~\cite{fireflies,crickets,WINFREE196715} for some examples. Interestingly,
achieving such mutual synchronization requires a cooperative effort, which is
undertaken without any central coordination between individuals. In
this
experiment we reproduce the behavior of pulse-coupled oscillators as described
in \cite{mirollo1990synchronization} and implemented in the \argos
simulation environment (see \Cref{fig:sync}). The model
involves a population of $n$
MarXbots~\cite{DBLP:conf/iros/BonaniLMRBRVBM10}, each of which is
equipped with an LED. For $i \in \{1, \ldots, n\}$, the $i$th robot is
characterized by a clock $x_i$ subject to the continuous dynamics
$\dot{x}_i = 1$, which applies as long as $0 \leq x_i \leq f$ for some
\textit{firing threshold} $f\in\mathbb{R}_{>0}$. When $x_i= f$, robot $i$ flashes its
central LED and $x_i$ is reset to zero by a discrete event. Robots
interact by a simple form of pulse coupling: when robot $i$ flashes,
all other robots are pulled towards firing according to the following
relation for some $\alpha\in\mathbb{R}_{>1}$:
\begin{equation}
	\label{eq:update}
	x_i = f \implies x_j := \left\{\begin{array}{ll}
		\alpha\cdot x_j & \text{ if } \alpha\cdot x_j < f \\
		0               & \text{ otherwise}
	\end{array}\right.
	\quad \textit{for all}\ j\in\{1,\ldots,n\}\setminus\{i\}
\end{equation}
Note that Eq.\ \ref{eq:update} only describes the update of the clocks -- neither the flashing nor implicit flashing of a robots' LED whenever the clock is reset to 0 is described. These properties have to be added to the model to make them observable.
Despite the simple model, the problem of pulse coupling represents a good example of how global swarm behaviors can emerge in distributed systems without being explicitly specified by individual control algorithms. Indeed, a global synchronization of flashing behaviors is achieved -- \ie, all clocks are synchronized -- even though this is not explicitly imposed by individual controllers.

\paragraph*{Modeling.} We propose several approaches on how to model the
synchronization problem compositionally. Each robot is modeled by a
hybrid automaton $\mathcal{H}_i$ such that the swarm behavior for a swarm of
size $n$ is modeled by a hybrid automaton $\mathcal{H}$ obtained by parallel
composition $\mathcal{H} = \mathcal{H}_1 || \ \ldots || \mathcal{H}_{n}$ of the
single components.
In the following we will discuss the different approaches in detail. All approaches model the case distinction of Eq. \ref{eq:update} via separate jumps to reflect the respective clock updates.

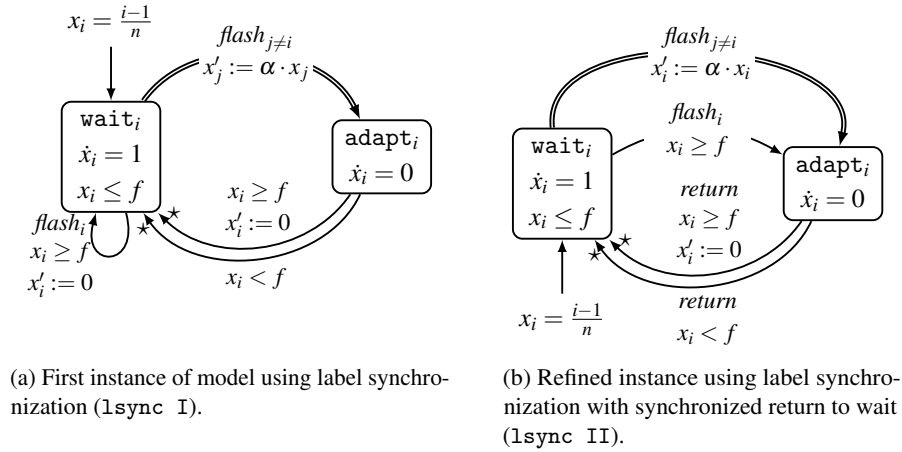
\begin{figure}
	\centering
	\subfloat[First instance of model using label synchronization (\lsync).\label{fig:labelSync}]{
		\raisebox{.04\textwidth}{\scalebox{0.9}{\begin{tikzpicture}[>=latex, state/.style={draw,rectangle,rounded corners,thick, text width=1.2cm, text centered}, every node/.style={text centered},node distance=4cm]
	\node[] (init) {$x_i = \frac{i-1}{n}$};
	\node[node distance=2cm, state,below of=init] (wait) {\texttt{wait}$_i$\\[.05cm]
										$\dot x_i = 1$\\[.05cm]
										$ x_i \leq f$};
	\node[state,right of=wait] (adapt) {\texttt{adapt}$_i$\\[.05cm]
										$\dot x_i = 0$};

	\draw[thick,->] (init) -- (wait);
	\draw[thick,->] (wait) edge[bend left=60,double] node[fill=white,inner sep=1pt, text width=1.6cm] {\small$\mathit{flash}_{j\neq i}$\\$x_j':=\alpha\cdot x_j$} (adapt);
	\draw[thick,->] (adapt) edge[bend left=50] node[above,text width=1cm] {\small$x_i \geq f$\\$x_i':=0$} node[pos=1,right]{$\star$}(wait);
	\draw[thick,->] (adapt) edge[bend left=60] node[below,text width=1cm] {\small$x_i < f$} node[pos=1,below]{$\star$} (wait);
	\draw[thick,->] (wait) edge[loop below,looseness=5] node[pos=.5,text width=1cm,left,xshift=-.1cm] {\small$\mathit{flash}_i$\\$x_i \geq f$\\$x_i':=0$} (wait);
\end{tikzpicture}}}
	}
	\qquad
	\subfloat[Refined instance using label synchronization with synchronized return to wait (\lsyncOpt).\label{fig:labelSyncOpt}]{
		\scalebox{0.9}{\begin{tikzpicture}[>=latex, state/.style={draw,rectangle,rounded corners,thick, text width=1.2cm, text centered}, every node/.style={text centered},node distance=4cm]
	\node[] (init) {$x_i = \frac{i-1}{n}$};
	\node[node distance=2cm, state,above of=init] (wait) {\texttt{wait}$_i$\\[.05cm]
										$\dot x_i = 1$\\[.05cm]
										$ x_i \leq f$};
	\node[state,right of=wait] (adapt) {\texttt{adapt}$_i$\\[.05cm]
										$\dot x_i = 0$};

	\draw[thick,->] (init) -- (wait);
	\draw[thick,double,->] (wait) edge[bend left=100,double] node[fill=white,inner sep=1pt, text width=1.5cm] {\small$\mathit{flash}_{j\neq i}$\\$x_i':=\alpha\cdot x_i$} (adapt);
	\draw[thick,->] (wait) edge[bend left=30] node[fill=white,inner sep=1pt, text width=1.5cm] {\small$\mathit{flash}_{i}$\\$x_i \geq f$
        } (adapt);
	\draw[thick,->] (adapt) edge[bend left=50] node[above,text width=1cm] {\small$\mathit{return}$\\$x_i \geq f$\\$x_i':=0$} node[pos=1,right]{$\star$}(wait);
	\draw[thick,->] (adapt) edge[bend left=60] node[below,text width=1cm] {\small$\mathit{return}$\\$x_i < f$} node[pos=1,below]{$\star$}(wait);

\end{tikzpicture}}
	}

	\caption{Models of one robot in the synchronization benchmark using label synchronization; jumps marked with a star are urgent, double arrows represent a set of synchronizing jumps, one for each $j\not=i$.\label{fig:HASync}}
\end{figure}

\parrep{Label synchronization.} The first hybrid automaton model for a single robot $r_i$  is shown in
\Cref{fig:labelSync}. It has two control modi $\texttt{wait}_i$ and $\texttt{adapt}_i$. To model the system using \emph{label synchronization}, we introduce synchonization labels $\mathit{flash}_i$, $i=1,\ldots,n$. Initially, the clocks $x_i$ of the robots $r_i$ start with different values in the location $\texttt{wait}_i$. When the clock valuation of robot $r_i$ reaches $f$,
in the respective automaton a transition with label $\mathit{flash}_i$ gets
enabled; the invariant assures that time cannot further evolve. If the jump with label $\mathit{flash}_i$ is taken, $r_i$ resets $x_i$ to $0$ and returns to its $\texttt{wait}_i$ mode, all other robots $r_j$, $j\not=i$ synchronize and take their jump with the label $\mathit{flash}_i$ simultaneously to their $\texttt{adapt}_j$ mode. Note that for each robot $r_j$, there is a jump from $\texttt{wait}_j$ to $\texttt{adapt}_j$ with label $\mathit{flash}_i$ for each $i\not=j$, we denote this in \Cref{fig:labelSync} by a double-lined arrow. Due to enabled urgent jumps back to $\texttt{wait}_j$, time cannot pass in $\texttt{adapt}_j$, but all robots $r_j$, $j\not=i$ will return using one of their two jumps, modeling the two cases in Eq. \ref{eq:update}.

While this model for a single component is relatively simple, using
a large number of synchronizing transitions has a strong impact on the size of the
parallel composition, as the number of transitions
drastically increases with the number of robots (see
\Cref{tab:automataStats}). Especially, after synchronizing on $\textit{flash}_i$, all robots $r_j$, $j\not=i$ will return from $\texttt{adapt}_j$ to $\texttt{wait}_j$ before time can further pass. However, since this returning is not synchronized, it can happen in all possible interleaving order, which adds unnecessary complexity. As an improvement, we could apply \emph{partial order reduction} by introducing a fixed order of execution for returning. However, in this special case we can even synchronize all
returning transitions,  which results in the improved model
\lsyncOpt shown in Figure \ref{fig:labelSyncOpt}.
In this model, we implement the desired behavior by making the returning jumps synchronized on the label \textit{return}.
Note that such simplifications need to be carefully designed to assure semantical equivalence, and are hard to automate. In our example, since robot $r_i$ will also have the label \textit{return}, it also needs to move to $\texttt{adapt}_i$ in order not to block the others, and will return using the jump with guard $x_i\geq f$.

\begin{figure}[t]
	\centering
	\subfloat[\label{fig:sharedVar}First model using shared variables (\shd).]
	{
		\scalebox{0.9}{\begin{tikzpicture}[>=latex, state/.style={draw,rectangle,rounded corners,thick, text width=1.2cm, text centered}, every node/.style={text centered},node distance=4cm]
	\node[text width=1.3cm, inner sep=0pt] (init) {$x_i = \frac{i-1}{n}$\\$z=0$};
	\node[node distance=1.7cm, state,right of=init] (wait) {\texttt{wait}$_i$\\[.05cm]
										$\dot x_i = 1$\\
										$\dot z = 0$\\[.05cm]
										$ x \leq f$};
	\node[state,above right of=wait] (adapt) {\texttt{adapt}$_i$\\[.05cm]
										$\dot x_i = 0$\\
										$\dot z = 0$};

	\node[state,below right of=wait,node distance=3cm] (flash) {\texttt{flash}$_i$\\[.05cm]
										$\dot x_i = 0$\\
										$\dot z = 0$};

	\draw[thick,->] (init) -- (wait);
	\draw[thick,->] (wait) edge[bend left] node[left, text width=2cm,xshift=-.2cm] {\small$\mathit{sync}_{1}$\\$x_i < f$\\$z = 1$\\$x_i':=\alpha\cdot x_i$} (adapt);
	\draw[thick,->] (adapt) edge[bend left=20] node[left,text width=1cm,yshift=.5cm,xshift=.2cm] {\small$\mathit{sync}_2$\\$x_i \geq f$\\$x_i':=0$} node[pos=.95,above]{$\star$}(wait);
	\draw[thick,->] (adapt) edge[bend left=30] node[right,text width=1cm,xshift=.1cm] {\small$\mathit{sync}_2$\\$x_i < f$}node[pos=.95,below]{$\star$} (wait);
	\draw[thick,->] (wait) edge[bend right=20] node[text width=2.2cm,left,pos=1, xshift=-0.3cm] {\small$x_i \geq f$\\$x_i':=0,z':=1$
	} (flash);
	\draw[thick,->] (flash) edge[loop right,looseness=5] node[right,text width=1cm]{\small$\mathit{sync}_1$\\$z = 1$\\$z':=0$} node[pos=.95,below]{$\star$}(flash);
	\draw[thick,->] (flash) edge[bend right=20] node[right,text width=1cm, xshift=.1cm]{$\mathit{sync}_2$\\$z = 0$} node[pos=.9,below]{$\star$} (wait);
\end{tikzpicture}}
	}
	\hfill
	\subfloat[\label{fig:sharedVarRed}Reduced instance using shared variables (\shdOpt).]
	{
		\scalebox{0.9}{\begin{tikzpicture}[>=latex, state/.style={draw,rectangle,rounded corners,thick, text width=1.2cm, text centered}, every node/.style={text centered},node distance=4.3cm]
	\node[text width=1.3cm] (init) {$x_i = \frac{i-1}{n}$\\$z=0$};
	\node[node distance=2cm, state,below of=init] (wait) {\texttt{wait}$_i$\\[.05cm]
										$\dot x_i = 1$
										$\dot z = 0$\\[.05cm]
										$ x \leq f$};
	\node[state,right of=wait] (adapt) {\texttt{adapt}$_i$\\[.05cm]
										$\dot x_i = 0$\\
										$\dot z = 0$};

	\draw[thick,->] (init) -- (wait);
	\draw[thick,->] (wait) edge[bend left=60] node[above, text width=2.5cm] {\small$\mathit{sync}_1$\\$z = 1 \land x_i < f$\\$x_i':=\alpha\cdot x_i, z':=0$} (adapt);
	\draw[thick,->] (wait) edge[bend left=45] node[below, text width=2.2cm] {\small$x = f$\\$x_i':=0,z':=1$} (adapt);

	\draw[thick,->] (adapt) edge[bend left=45] node[above,text width=2cm] {\small$\mathit{sync}_2$\\$x_i \geq f\land z = 0$\\$x_i':=0$} node[pos=.95,above]{$\star$}(wait);
	\draw[thick,->] (adapt) edge[bend left=60] node[below,text width=2cm] {\small$\mathit{sync}_2$\\$x_i < f \land z = 0$} node[pos=.95,below]{$\star$}(wait);
	\draw[thick,->] (adapt) edge[loop below,looseness=5] node[text width=1cm,below] {\small$\mathit{sync}_1$\\$z = 1$\\$z':=0$} node[pos=.95,right]{$\star$}(adapt);
\end{tikzpicture}}
	}

	\caption{Hybrid automata modelling a single robot in the synchronization benchmark using a shared variable $z$ for synchronization. Urgent transitions are marked with a star.}
\end{figure}
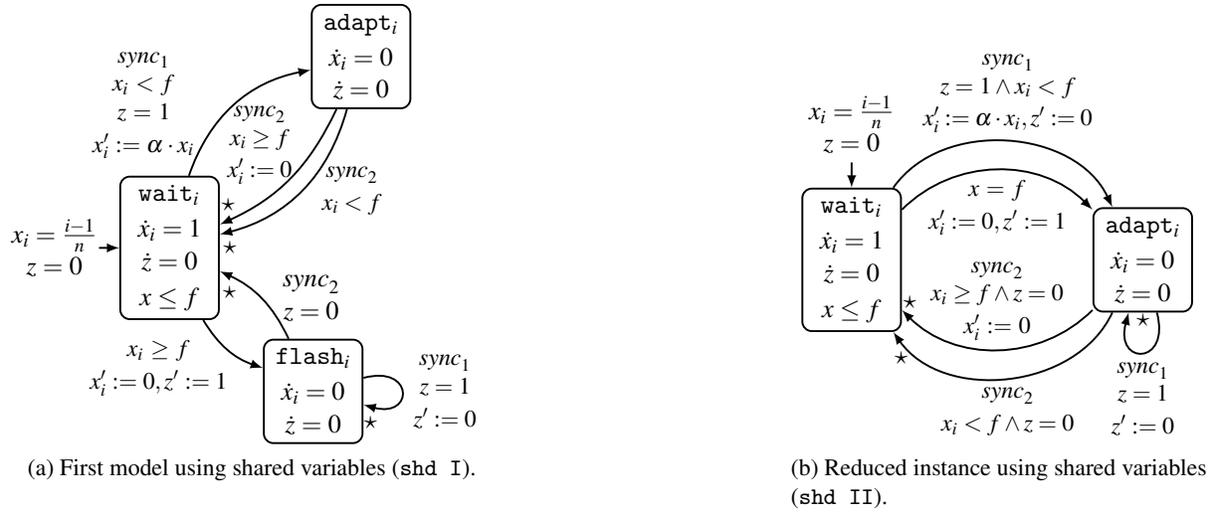

\parrep{Shared variables.} Our previous models have two main drawbacks: Firstly, they have a large number of jumps. Secondly, modeling a robot requires full information about the total number of components (and their respective synchronization labels), which does not allow for a generic approach.
To achieve a model with less jumps and where single components do not require any prior knowledge about the full system, we make use of \emph{shared variables}. The idea is not to distinguish on which of the robots flashes, but use a shared flag $z$ to model the fact that \emph{at least one} of the robots flashes. Our first model using shared variables is shown in \Cref{fig:sharedVar}. Starting with $z=0$ initially, once the clock $x_i$ of robot $r_i$ reaches the threshold $f$ it flashes, which is modeled by a jump from $\texttt{wait}_i$ to a new location $\texttt{flash}_i$, which sets $z$ to $1$ to model that a flashing took place. This location has an enabled urgent jump, thus time cannot proceed further. All other robots that flash at the same time need to move to their $\texttt{flash}$ mode first, in order to be able to synchronize among all the robots on the label $\textit{sync}_1$; the execution of these synchronized steps adapt the clocks of the non-flashing robots (that move to their $\texttt{adapt}$ mode) and set $z$ back to $0$ (on the self-loop of the $\texttt{flash}_i$ mode). A second step synchronizes again all robots, this time on the label $\textit{sync}_2$, to get back to their wait mode.

Note how the use of a shared variable allowed us to abstract away from the identity of the flashing robot(s): instead of introducing an individual synchronization for the flashing of each of the robots, any flashing robot can set the flag $z$ and trigger the same synchronization process. Note furthermore that this flag is indeed needed to correctly implement this abstraction: when we would remove it, all robots could adapt (move from waiting to adapting) without any flash happening. Finally, observe that this shared flag allowed to strongly reduce the number of jumps locally in the components as well as globally in the composition, as we do not need to distinguish on the flashing robot's identity any more.

A reduced version \shdOpt, shown in \Cref{fig:sharedVarRed}, uses a similar mechanism but unifies the locations $\texttt{adapt}_i$ for adaption and $\texttt{flash}_i$ for setting the synchronization flag into one location to reduce the parallel composition's size.

\begin{table}[t]
	\caption{\label{tab:automataStats}Number of locations (\#locs.) and transitions (\#trans.) in the resulting automata for different numbers of robots (\#robots).}
	\centering
	\begin{tabularx}{\textwidth}{llXXXXXXXX}
		\toprule
		\multicolumn{1}{c}{}      & \multicolumn{1}{c}{} & \multicolumn{8}{c}{\#robots}                                              \\
		                          & version              & 1                            & 2  & 3  & 4   & 5   & 6    & 7     & 8     \\\midrule
		\multirow{3}{*}{\#locs.}
		                          & \lsync               & 2                            & 3  & 7  & 15  & 31  & 63   & 127   & 255   \\
		& \lsyncOpt            & 2                            & 3  & 4  & 5   & 6   & 7    & 8     & 9     \\
                & \shd                 & 3                            & 7  & 15 & 31  & 63  & 127  & 255   & 511   \\
		                          & \shdOpt              & 2                            & 4  & 8  & 16  & 32  & 64   & 128   & 256   \\
		\midrule
		\multirow{3}{*}{\#trans.}
		                          & \lsync               & 3                            & 6  & 33 & 164 & 755 & 3310 & 14077 & 58728 \\
		& \lsyncOpt            & 3                            & 6  & 15 & 36  & 85  & 198  & 455   & 1032  \\
                & \shd                 & 6                            & 18 & 54 & 162 & 486 & 1458 & 4374  & 13122 \\
		                          & \shdOpt              & 5                            & 13 & 35 & 97  & 275 & 793  & 2315  & 6817  \\
		\bottomrule
	\end{tabularx}
\end{table}

To compare our approaches we have created models for the synchronization benchmark for
up to $8$ robots using all presented approaches. Statistics about the resulting hybrid
automata are listed in \Cref{tab:automataStats}.
From our results we can observe, that the natural approach via label synchronization (\lsync) creates a high number of transitions in the resulting composed automaton while keeping the number of locations low. Applying the optimization (\lsyncOpt) which effectively collects sequences of urgent transitions into a single urgent transition, reduces the number of transitions drastically.
The versions using shared variables (\shd~+ \texttt{II}) produce results with
less transitions, \eg, edges for the return to \texttt{wait} are collected by a
synchronization label ($\mathit{sync}_2$) and thus create one jump in the
parallel composition.

\begin{figure}
	\centering
	\includegraphics[scale=.53]{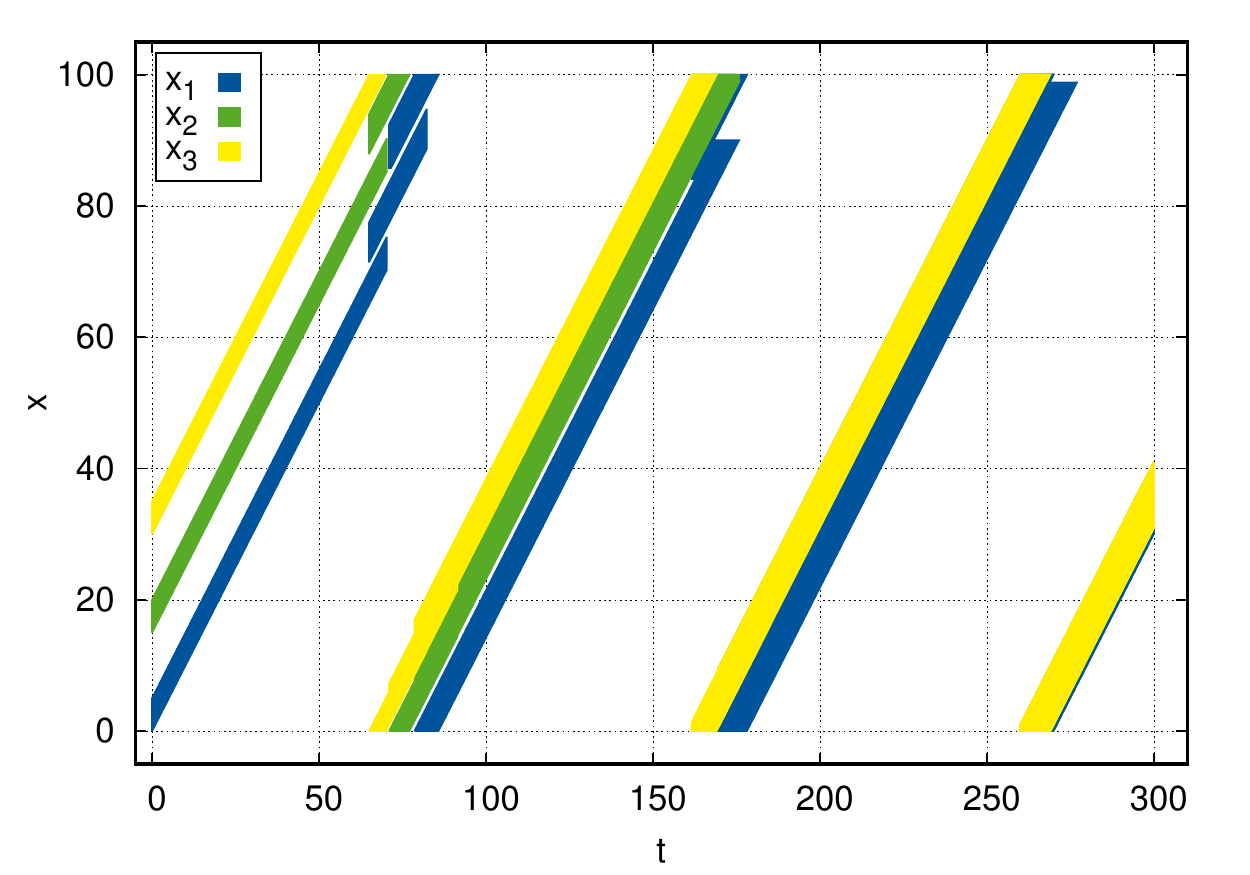}
	\includegraphics[scale=.53]{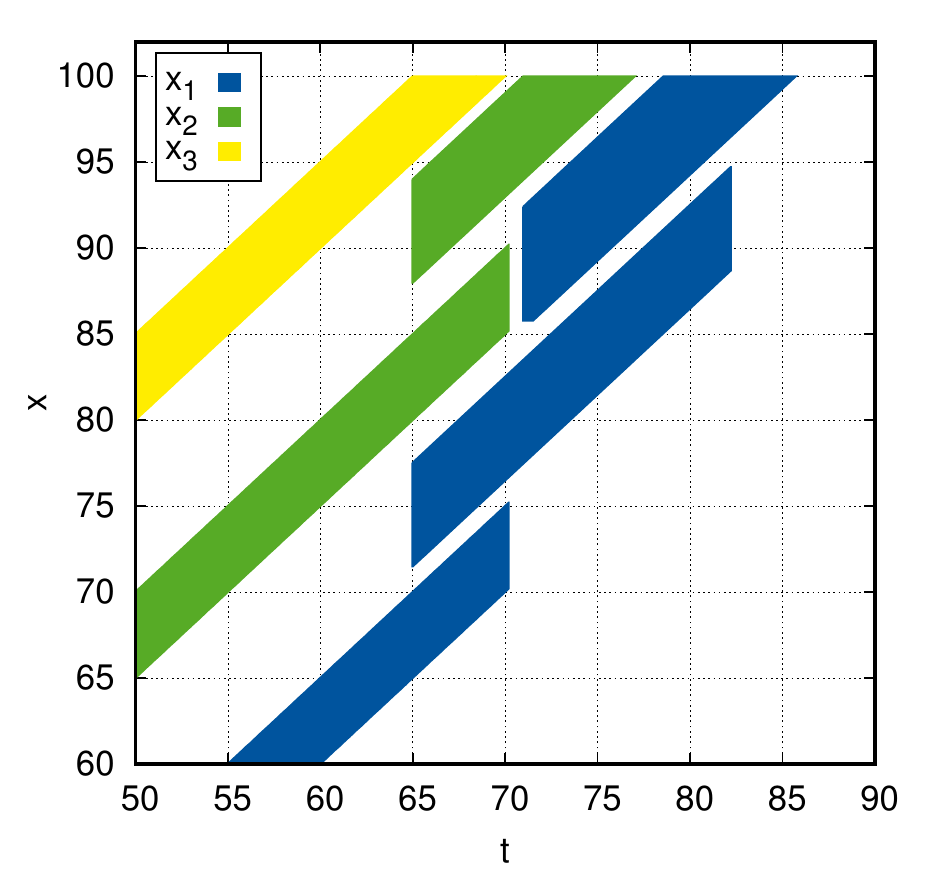}
	\caption{Overlayed flowpipes for a system of 3 robots for $f=100, \alpha=1.1$ (local time horizon 110 sec.). Right: excerpt showing clock adaption after flashing.\label{fig:experimentFlowpipes}}
\end{figure}

\paragraph*{Reachability analysis results.}
\Cref{fig:experimentFlowpipes} shows the flowpipes for a system with three robots and parameters chosen to illustrate how the synchronization occurs and that the analysis is working with sets of states rather than single trajectories. The right image emphasizes a phenomenon that arises from the latter fact.

In \Cref{tab:runningTimes} the running times for our experiments with different
numbers of robots using the different modeling approaches can be found. The
initial clock valuations for robots $r_i$ were chosen as $\frac{i-1}{n}$ as
indicated in the models.

\bgroup
{\setlength{\tabcolsep}{10pt}
	\begin{table}
		\caption{\label{tab:runningTimes}Running times for different numbers of robots using $f=1, \alpha=1.1$ (time step size: $\delta=0.01$, jump depth: $20$, state set representation: boxes). Timeout (TO) is $2$ minutes.}
		\centering
		\begin{tabularx}{\textwidth}{lXXXXXXXX}
			\toprule
			\multicolumn{1}{c}{} & \multicolumn{8}{c}{\# robots}                                                                                                     \\
			version              & 1                             & 2             & 3             & 4             & 5             & 6             & 7            & 8  \\\midrule
			\lsync               & 0.12                          & \textbf{0.11} & 0.13          & 0.21          & 1.02          & 64.6          & TO           & TO \\
			\lsyncOpt            & 0.12                          & \textbf{0.11} & \textbf{0.12} & \textbf{0.14} & \textbf{0.25} & \textbf{2.96} & \textbf{146} & TO \\
			\shd                 & \textbf{0.11}                 & \textbf{0.11} & \textbf{0.12} & 0.16          & 0.48          & 9.88          & TO           & TO \\
			\shdOpt              & \textbf{0.11}                 & \textbf{0.11} & 0.13          & 0.19          & 0.75          & 16.85         & TO           & TO \\
			\bottomrule
		\end{tabularx}
	\end{table}
}
\egroup

\newcommand{\setEnum}[1]{\{#1\}}
\newcommand{\set}[2]{ \setEnum{#1 \mid #2}}
\newcommand{\sets}[2]{\setEnum{#1, \hdots, #2}}

\subsection{Scalability Improvements}

Our results so far are unsatisfactory in two ways:
first, the number of robots is limited to at most seven and second, the jump depth limit of 20 does not allow to verify that synchronization across all robots occurs.
In Sections~\ref{subsubsec:compAnalysis} and \ref{subsubsec:EdgeOpt} we apply two optimizations to increase the number of robots the analysis can handle and then in Section~\ref{subsubsec:explicitTime} we optimize precision to allow us to increase jump depth and verify synchronization for a limited number of robots.

We focus on the \lsyncOpt model because with the label synchronization
approach, synchronization is limited to discrete jumps via labels which we believe increases our chances of success.
Focusing on this model allows for several improvements as we show in the next sections.
We briefly describe the techniques we use and show intermediate results to illustrate the resulting effect.
Each technique is independent from the others and is generalizable for the analysis of any linear hybrid automaton, although the effectiveness depends on the structure of the input.
In the following experiments, first we have used unique initial states to examine scalability and to analyze the source of imprecision; in Section \ref{subsubsec:explicitTime} we will extend our experiments to sets of initial states.



\subsubsection{Compositional Analysis}
\label{subsubsec:compAnalysis}

Instead of computing the parallel composition of the input automata for each robot eagerly, our aim is to perform an on-the-fly, decomposed analysis.
We start with only the initial states and then lazily compute the reachable parts of the automaton as we discover them.
This helps to cope with the size of the resulting automaton, as we do not construct locations that are not reached within the bounded analysis and only construct transitions from reachable locations.

We also compute the continuous successors of each automaton individually, i.e., a flowpipe for each automaton where each flowpipe consists of state sets over only the variables of the corresponding automaton, similar to our previous approach in~\cite{schupp2017divide}.
In the specific case of the synchronization problem, we obtain $n$ flowpipes
as intervals, since each automaton has only one variable.
This reduces the dimension of the state sets, but also reduces precision as we will see next.

\paragraph*{Precision.}
Since we separate the variables of the component automata and operate in separate sub-spaces, we lose information on the relation between the variables.
This can be illustrated by the following example.

Assume we are using a representation that can exactly describe the trajectories of linear functions, such as oriented boxes or polytopes.
Given two robots from our synchronization problem modeled as in \lsyncOpt, using a single composed state space built from variables $x_1, x_2$ and starting with initial valuation $(0, \frac{1}{2})\subseteq\R^2$.
After taking a time step of $\delta=\frac{1}{2}$ all reachable valuations are contained in the set $\set{(x, x + \frac{1}{2})}{0\leq x \leq \frac{1}{2}}$.
Intersecting this set with $x_2 \geq 1$ results in the singleton set $\{(\frac{1}{2}, 1)\}$.
In contrast to this, when working with separate state spaces, we obtain $[0,\frac{1}{2}] \times [\frac{1}{2}, 1]$ as the set of reachable valuations.
Due to the missing relation between the separate state spaces (in this case the relation $x_2 = x_1 + \frac{1}{2}$), the intersection with $x_2 \geq 1$ gives $[0, \frac{1}{2}] \times [1,1]$.

Note that here the resulting state sets are the same as when using a box representation in the original, composed state space, but the approach does not allow for the sensible use of more precise representations in our case since the problem is one-dimensional in each sub-space.
In higher-dimensional sub-spaces, more precise state set representations may improve precision within each subspace but do not affect the loss of information when synchronizing sub-spaces via Cartesian product.

\paragraph*{Results.}
\begin{table}[t]
  \caption{\label{tab:robotsCompositional13} Running time in seconds and number of nodes in reachability tree for $n$ robots with $f=1$, $\alpha= 1.3$, time step size $\delta=
    10^{-5}$ and max jump depth of $20$. Transition enumeration is not optimized. Deduplication (see \Cref{subsubsec:explicitTime}) is used. Timeout (TO) is $2$ minutes.}
    \centering
	\begin{tabularx}{\textwidth}{l X X X  X  X  X  X  X  X  X  X X X X X X }
    \toprule
    $n=$ & $1$ & $3$ & $5$ & $7$ & $9$ & $11$ & $13$ & $15$ & $17$ & $19$ & $21$ & $23$ & $25$ \\
    \midrule
    time & 5.21 & 3.56 & 4.08 & 6.69 & 7.13 & 6.43 & 15.5 & 13.0 & 9.54 & 8.51 & 27.3 & 36.7 & TO \\
    nodes & 21 & 21 & 21 & 21 & 21 & 21 & 40 & 30 & 21 & 21 & 28 & 21 & NA \\
    \bottomrule
    \end{tabularx}
\end{table}

In Table~\ref{tab:robotsCompositional13} the running times and number of nodes in the generated reachability tree for different swarm sizes are shown.
The loss of precision that we previously discussed can contribute to branching in the reachability tree as this potentially allows for discrete non-determinism when several transitions are enabled.
More concretely, if there are reachable state sets with a clock value below and above $f$ in the location \texttt{adapt}, then successors of both return transitions must be considered, which causes exponential branching behavior.
To circumvent this, we use a very small time step of $\delta=10^{-5}$---nonetheless some branching still occurs, which is indicated by node numbers larger than $21$.

The compositional approach has already improved the scalability of the
analysis, but an exponential number of transitions is still enumerated, causing
running times to increase beyond feasibility for larger time horizons or more
robots, even when using small initial state sets or even point-sets.
Next we look at a way to avoid this exponential explosion.

\subsubsection{Optimized Transition Enumeration}
\label{subsubsec:EdgeOpt}

Until now, scalability has been impeded by the exponential number of transitions leading from the composed \texttt{adapt} location to the \texttt{wait} location.
Each of the $n$ automata may take either of its two return transitions, which leads to $2^n$ transitions in the composed system.
However, when considering the actual reachable states of the automata, we observe that in almost all cases only one of the transitions is enabled for each automaton.
As we are constructing the jump successors lazily, i.e., we construct them after computing the continuous successors in form of a flowpipe, the information needed to check which transitions are enabled is available to us.

To use this information we chose an arbitrary, but fixed order for the automata.
Based on this order, we iteratively determine for each of the transitions (in our example the two return transitions) which flowpipe segments enable the jump in the current automaton.
For each transition with one or more enabling segments, we then check for the next automaton which transitions the corresponding segments (synchronized by time) of its flowpipe enable to iteratively restrict the number of segments enabling the  considered jump.
We enumerate all combinations of transitions with enabling segments in all automata, but do not consider combinations for which there is not a segment enabling the respective jump in each automaton.

In our case this means that we construct exactly one transition from the composed \texttt{adapt} location to the composed \texttt{wait} location, unless there is branching due to loss of precision, as mentioned above.

\paragraph*{Results.}
\begin{table}[t]
  \caption{\label{tab:robotsCompositionalOpt13} Running time in seconds and number of nodes in reachability tree for $n$ robots with $f=1$, $\alpha= 1.3$, $\delta=
    10^{-5}$ and max jump depth of $20$. Transition enumeration is optimized. Deduplication (see \Cref{subsubsec:explicitTime}) is used. }
    \centering
	\begin{tabularx}{\textwidth}{l X X X  X  X  X X  X  X  X X XX X X X  }
    \toprule
    $n=$ & $25$ & $50$ & $75$ & $100$ & $125$ & $150$ & $175$ & $200$ & $225$ & $250$ & $275$ & $300$ & $325$ & $350$ & $375$ & $400$ \\
    \midrule
    time & 13.6 & 3.67 & 4.18 & 3.97 & 4.04 & 6.34 & 7.13 & 13.6 & 4.44 & 5.16 & 9.54 & 19.0 & 19.5 & 23.3 & 12.5 & 33.1 \\
    nodes & 21 & 22 & 22 & 22 & 22 & 27 & 30 & 22 & 22 & 22 & 41 & 40 & 40 & 46 & 30 & 26 \\
    \bottomrule
    \end{tabularx}
\end{table}

In Table~\ref{tab:robotsCompositionalOpt13} we see the results of the same experiment as before, but with the transition enumeration optimization applied.
Clearly, the scalability has greatly improved.
Notably, the branching behavior has not fundamentally changed and the maximal jump depth is still set to 20, meaning we can not verify synchronization.
Increasing the jump depth leads to more branching because more jumps are made and also because precision keeps decreasing with every jump as over-approximation errors accumulate.

\subsubsection{Explicit Time Dimension and Successor Deduplication}
\label{subsubsec:explicitTime}
To recover the information on the correlation of variable valuations in the different automata, we introduce a variable in each automaton that tracks the time, i.e., its initial value is zero and its derivative is one.
When determining the sets of states that enable a transition on which multiple automata synchronize, which is true for all transitions in our example, we know that the time at which the transition is taken must be the same in all automata.
We use this by projecting the enabling sets of all synchronizing automata on their time dimension to obtain a time interval during which the transition is enabled.
We then intersect these intervals, since all automata must be allowed to jump at the same time, and intersect the enabling sets of the automata with the resulting time interval.
For this to be effective, we use a representation that can precisely describe the relation between time and each automaton's own variable---here we choose template polyhedra with an octagonal template (i.e., polyhedra in half-space representation where the normal vectors are fixed).

For our example from Section~\ref{subsubsec:compAnalysis} this means that for the automaton starting at $x_1=\frac{1}{2}$, we obtain the point interval $[\frac{1}{2}, \frac{1}{2}]$ for the time during which the transition with guard $x\geq 1$ is enabled.
Intersecting the other automaton's enabling segment with this gives us the point interval $[\frac{1}{2}, \frac{1}{2}]$ for its value of $x_2$, i.e., we are able to fully recover the lost information on the relation of the variables.

\paragraph*{Successor deduplication.}
We can now increase the jump depth to verify the synchronization of the robots' clocks.
But the non-determinism that occurs when the clocks do actually synchronize still limits scalability:
Since multiple robots may flash at the exact same time and the transition synchronizes on the flash-event of a specific robot $i$, each of the automata can take the transition with label $\mathit{flash}_i$, i.e., represents the robot that triggers the flash, or ``follows'' another robot via the transition labeled $\mathit{flash}_{j\neq i}$.

However, it effectively does not matter which automaton takes which transition, since they all end up in \texttt{adapt} and their valuations are all modified in the same way.
Cases like this can be identified by identical resets on (synchronized)
transitions, i.e., several transitions with the same reset function.
Note that while this is generalizable, it makes sense to combine transitions in which the enabling segments overlap---otherwise the approach will introduce additional over-approximation.

In our example, since all those transitions are taken during the same time interval and all of those transitions apply the same reset function, applying the reset function on the union of segments enabling the transitions has the same effect as applying it individually on each enabling segment.
As a result, we can resolve the ambiguity between robots which flash simultaneously and produce only one discrete successor which significantly reduces the branching in the resulting reachability tree.

\paragraph*{Results.}
\begin{table}[t]
  \caption{\label{tab:robotsCompositionalOptEtDd} Running time in seconds and number of nodes in reachability tree for $n$ robots with $f=1$, $\alpha= 1.3$, $\delta=0.01$ and termination once synchronization was detected. We used transition enumeration optimization, transition deduplication and an explicit time dimension.}
    \centering
	\begin{tabularx}{\textwidth}{l X X X  X  X  X  X  X  X  X  X X X X X X }
    \toprule
    $n=$ & $40$ & $80$ & $120$ & $160$ & $200$ & $240$ & $280$ \\
    \midrule
    time & 24.08 & 47.79 & 81.56 & 80.45 & 104.1 & 112.4 & 232.1 \\
    nodes & 52 & 54 & 58 & 48 & 56 & 52 & 58 \\
    \bottomrule
    \end{tabularx}
\end{table}

In Table \ref{tab:robotsCompositionalOptEtDd} we have changed the time step, as it does not need to be as small as before to counteract imprecision.
We also removed the cap on jumps and instead terminate when we detect that all automata are synchronized.
Thus the number of nodes corresponds to twice the number of flashes needed to
establish synchronization, as each flash requires taking two jumps (from
\texttt{wait} to \texttt{adapt} and back).

It is clear that the more sophisticated representation of template polyhedra
and overhead of projections increase running times by a noticeable factor, but
the desired property of clock synchronization can still be shown for large
swarms in reasonable time.

Note that until now we have used point-sets for each clock of each automaton as our first goal was to increase scalability of the general reachability analysis for composed systems and to understand the reasons for the loss of precision and exponential blow-up caused by compositional analysis.
As we could show synchronization for these cases after adding several features based on our observations, we next ran our experiments with a setup where the clock valuations of the automata are taken from intervals.
The results for different widths of initial sets and different values for $\alpha$ are shown in \Cref{tab:robotsCompositionalOptIntervals} using breadth-first search and in \Cref{tab:robotsCompositionalOptIntervalsDFS} using depth-first search.

We can see that while our previous improvements seemed promising in the sense that they reduced the exponential blow-up in discrete jumps caused by the compositional analysis, the introduced non-determinism from reasonably-sized initial sets still poses a large problem when trying to prove synchronization via flowpipe-construction-based reachability analysis.
From the tables we can see, that the non-determinism results in large search trees, which render the analysis infeasible---the running time for single nodes seems negligible, since runs where the search tree has a reasonable size still terminate in less than a second.

\begin{table}[t]
  \caption{\label{tab:robotsCompositionalOptIntervals} Running time in seconds and number of nodes in reachability tree for $1-3$ robots with $f=1$, time step size $\delta=
    10^{-5}$ and max jump depth of $10000$ using breadth-first search. Transition enumeration is optimized. Deduplication (see \Cref{subsubsec:explicitTime}) is used. }
    \centering
	\begin{tabularx}{\textwidth}{l XXXX XXXX XXXX}
    \toprule
    robots & \multicolumn{4}{c}{1} & \multicolumn{4}{c}{2} & \multicolumn{4}{c}{3} \\
    width & ver. & dpth. & nodes & time & ver. & dpth. & nodes & time & ver. & dpth. & nodes & time \\
    \midrule
    $\alpha= 1.3$ &&&&&&&&&&&& \\
	1/100 & 1 & 1 & 2 & 0 & 1 & 17 & 18 & 0 & 0 & 40 & 12721 & 60 \\
    1/50 & 1 & 1 & 2 & 0 & 1 & 17 & 26 & 0 & 0 & 38 & 12899 & 60 \\
    3/100 & 1 & 1 & 2 & 0 & 1 & 21 & 54 & 0 & 0 & 32 & 14179 &60 \\
    \midrule
    $\alpha= 1.2$ &&&&&&&&&&&& \\
    1/100 & 1 & 1 & 2 & 0 & 1 & 29 & 30 & 0 & 0 & 70 & 11095 & 60 \\
    1/50 & 1 & 1 & 2 & 0 & 1 & 33 & 66 & 0 & 0 & 46 & 12050 & 60 \\
    3/100 & 1 & 1 & 2 & 0 & 1 & 37 & 230 & 0 & 0 & 34 & 13468 &60 \\
    \midrule
    $\alpha= 1.1$ &&&&&&&&&&&& \\
    1/100 & 1 & 1 & 2 & 0 & 1 & 77 & 192 & 0 & 0 & 68 & 10625 & 60 \\
    1/50 & 1 & 1 & 2 & 0 & 1 & 101 & 1055 & 0 & 0 & 55 & 11022 & 60 \\
    3/100 & 1 & 1 & 2 & 0 & 1 & 120 & 13918 & 0 & 0 & 49 & 12046 &60 \\
    \bottomrule
    \end{tabularx}
\end{table}

\begin{table}[t]
  \caption{\label{tab:robotsCompositionalOptIntervalsDFS} Running time in seconds and number of nodes in reachability tree for $1-3$ robots with $f=1$, time step size $\delta=
    10^{-5}$ and max jump depth of $10000$ using breadth-first search. Transition enumeration is optimized. Deduplication (see \Cref{subsubsec:explicitTime}) is used. }
    \centering
	\begin{tabularx}{\textwidth}{l XXXX XXXX XXXX}
    \toprule
    robots & \multicolumn{4}{c}{1} & \multicolumn{4}{c}{2} & \multicolumn{4}{c}{3} \\
    width & ver. & dpth. & nodes & time & ver. & dpth. & nodes & time & ver. & dpth. & nodes & time \\
    \midrule
    $\alpha= 1.3$ &&&&&&&&&&&& \\
	1/100 & 1 & 1 & 2 & 0 & 1 & 17 & 18 & 0 & 0 & 161 & 9144 & 60 \\
    1/50 & 1 & 1 & 2 & 0 & 1 & 17 & 26 & 0 & 0 & 137 & 9252 & 60 \\
    3/100 & 1 & 1 & 2 & 0 & 1 & 21 & 54 & 0 & 0 & 221 & 9357 &60 \\
    \midrule
    $\alpha= 1.2$ &&&&&&&&&&&& \\
    1/100 & 1 & 1 & 2 & 0 & 1 & 29 & 30 & 0 & 0 & 289 & 8803 & 60 \\
    1/50 & 1 & 1 & 2 & 0 & 1 & 33 & 66 & 0 & 0 & 231 & 9155 & 60 \\
    3/100 & 1 & 1 & 2 & 0 & 1 & 37 & 230 & 0 & 0 & 211 & 8883 &60 \\
    \midrule
    $\alpha= 1.1$ &&&&&&&&&&&& \\
    1/100 & 1 & 1 & 2 & 0 & 1 & 77 & 192 & 0 & 0 & 113 & 7853 & 60 \\
    1/50 & 1 & 1 & 2 & 0 & 1 & 101 & 1055 & 0 & 0 & 133 & 7562 & 60 \\
    3/100 & 1 & 1 & 2 & 0 & 1 & 341 & 11436 & 0 & 0 & 111 & 7258 & 60 \\
    \bottomrule
    \end{tabularx}
\end{table}

%
%
%
%

\section{Conclusion}
\label{sec:future}
In this paper we have shown how algorithms and tools for the verification
of hybrid systems could be employed to analyze controllers for swarm
robotics.
Reachability analysis via flowpipe-construction was used as
the basis for the verification of global swarm behavior.
Experimental results show the potential of our approach, nevertheless, several challenges are yet to be addressed in order to
increase the applicability of such technique to robotics.
Modelling proved to be a challenging task: as observed in our case study, even for a simple system there are manifold ways to create a model which reflects its behavior.
While theoretically equivalent, our results show a strong impact of the modeling on the scalability of the analysis of the resulting model.
In this context, we have shown several improvements which can be generalized to other problems which allow to partially overcome the scalability issues related to compositional models.
Our experimental evaluation indicates the strong effect of those optimizations in a well-controlled environment and even allow to prove properties in some cases.
However, our results also indicate, that there is room for improvement, especially when we want to go beyond computing reachability in an over-approximative way but instead want to use results to prove properties for setups which are less controlled (i.e., exhibit more non-determinism, here reflected by larger initial configurations).
We hope this work will stimulate further investigations in this exciting and uncharted direction.

\paragraph*{Future work.}
Based on our case study we have shown approaches to improve scalability for compositional models. 
Several ways of further improving our method can be foreseen.

One direction for future work aims at implementing the lazy construction of locations and transitions while still using a single unified state space or potentially even clustering automata into several state spaces. 
This would allow us to effectively use precise representations while still dealing well with a large number of locations and transitions.
We see that the optimizations as shown in Section~\ref{subsubsec:EdgeOpt} can be generalized also for non-decomposed state spaces in compositional models which may represent a middle-ground between running time and precision for composed models using label synchronization.
Furthermore, in this case study we solely focused on label synchronization while shared variables were neglected.
Investigating dedicated techniques in this area may be another interesting thread of research, for instance, handling shared discrete variables separately may already prove useful.

\bibliographystyle{eptcs}
\bibliography{literature}




\end{document}